\documentclass[14pt]{article}
\usepackage{arxiv}
\usepackage[numbers,square,sort&compress]{natbib}
\usepackage{hyperref}
\usepackage{ulem}
\usepackage{booktabs}
\usepackage{xspace}
\usepackage{graphicx}
\usepackage{fancyhdr}
\usepackage{pslatex}   
\usepackage{xcolor}
\usepackage{bm}
\usepackage{color}
\usepackage{amsmath, amsthm, amssymb} 
\usepackage{amssymb}
\usepackage{enumerate}
\usepackage{enumitem}
\usepackage{subfigure}
\usepackage[symbol]{footmisc}

\usepackage{lineno}
\usepackage[utf8]{inputenc} 
\usepackage[T1]{fontenc}    

\usepackage{indentfirst} 
\usepackage{titlesec}
\titlespacing{\section}{0pt}{0\parskip}{-0.2\parskip}
\titlespacing{\subsection}{0pt}{0\parskip}{-0.5\parskip}

\newcommand{\gev}{\,\mathrm{GeV}}
\newcommand{\modea}{pK_{S}^0}
\newcommand{\modeb}{pK^{-}\pi^+}
\newcommand{\modec}{pK_{S}^0\pi^0}
\newcommand{\moded}{pK_{S}^0\pi^+\pi^-}

\newcommand{\modeaa}{\Lambda\pi^+}
\newcommand{\modebb}{\Lambda\pi^+\pi^0}
\newcommand{\modedd}{\Lambda\pi^+\pi^-\pi^+}
\newcommand{\modeaaa}{\Sigma^{0}\pi^+}
\newcommand{\modeccc}{\Sigma^{+}\pi^0}
\newcommand{\modeddd}{\Sigma^{+}\pi^+\pi^-}
\newcommand{\Modea}{\bar{p}{}K_{S}^0}
\newcommand{\Modeb}{\bar{p}{}K^{+}\pi^-}
\newcommand{\Modec}{\bar{p}{}K_{S}^0\pi^0}
\newcommand{\Moded}{\bar{p}{}K_{S}^0\pi^-\pi^+}

\newcommand{\Modeaa}{\bar{\Lambda}{}\pi^-}
\newcommand{\Modebb}{\bar{\Lambda}{}\pi^-\pi^0}
\newcommand{\Modedd}{\bar{\Lambda}{}\pi^-\pi^+\pi^-}
\newcommand{\Modeaaa}{\bar{\Sigma}{}^{0}\pi^-}
\newcommand{\Modeccc}{\bar{\Sigma}{}^{-}\pi^0}
\newcommand{\Modeddd}{\bar{\Sigma}{}^{-}\pi^-\pi^+}
\newcommand{\lambdacp}{\Lambda_{c}^{+}}
\newcommand{\lambdacm}{\bar{\Lambda}{}_{c}^{-}}

\title{Observation of a rare beta decay of the charmed baryon with a Graph Neural Network}
\author{\Large The BESIII Collaboration$^1$\\
$^1$A list of authors and their affiliations appears at the end of the preprint.\\}

\begin{document}
\maketitle
\begin{abstract}  
The beta decay of the lightest charmed baryon $\Lambda_c^+$ provides unique insights into the fundamental mechanism of strong and electro-weak interactions, serving as a testbed for investigating non-perturbative quantum chromodynamics and constraining the Cabibbo-Kobayashi-Maskawa (CKM) matrix parameters. This article presents the first observation of the Cabibbo-suppressed decay $\Lambda_c^+ \rightarrow n e^+ \nu_{e}$, utilizing $4.5~\mathrm{fb}^{-1}$ of electron-positron annihilation data collected with the BESIII detector. A novel Graph Neural Network based technique effectively separates signals from dominant backgrounds, notably $\Lambda_c^+ \rightarrow \Lambda e^+ \nu_{e}$, achieving a statistical significance exceeding $10\sigma$. The absolute branching fraction is measured to be $(3.57\pm0.34_{\mathrm{stat.}}\pm0.14_{\mathrm{syst.}})\times 10^{-3}$. For the first time, the CKM matrix element $\left|V_{cd}\right|$ is extracted via a charmed baryon decay as $0.208\pm0.011_{\rm exp.}\pm0.007_{\rm LQCD}\pm0.001_{\tau_{\Lambda_c^+}}$. This work highlights a new approach to further understand fundamental interactions in the charmed baryon sector, and showcases the power of modern machine learning techniques in experimental high-energy physics.
\end{abstract}
\clearpage

\section*{Introduction}
Beta decay, a natural radioactivity discovered in the early 20th century, opened a window to probe the subatomic matter world. Ernest Rutherford's observations in 1899~\cite{rutherford1899viii} initiated its recognition, followed by the elucidation of its complexities by Enrico Fermi in the 1930s~\cite{Fermi:1934sk}. This decay mechanism allows an atomic nucleus to transform into an isobar of that nuclide by emission of an electron (positron) and an anti-neutrino (neutrino). It involves certain intrinsic properties of subatomic particles and their interaction via the weak force, one of the fundamental interactions in nature. Beta decay exists in two types: first, a free (or bound) neutron may transform into a proton, an electron, and an anti-neutrino in $\beta^-$ decay $n\to p^+ e^{-} \Bar{\nu_{e}}$. Conversely, a bound proton within an unstable nucleus transforms into a neutron, a positron and a neutrino via $\beta^+$ decay $p^+\to n e^{+} {\nu_{e}}$. As with the bound neutron case, this $\beta^+$ decay happens only inside nuclei when the daughter nucleus has a sufficiently greater binding energy than the mother nucleus. Studying $\beta^+$ decay provides insights into the interactions between protons and neutrons within nuclei, revealing a complex interplay of gluons and quarks through the strong interactions which remains incompletely understood.  Scientists can gain complementary insights by studying analogous decays of $\Lambda$-type baryons, which present distinctive opportunities to study $\beta^+$ decay. These baryons are similar to neutrons and protons, but with the  replacement of a light quark with a heavy quark.  Examples, with the quark structure listed in parentheses, include $\Lambda(uds)$, $\Lambda_c^+(udc)$, and $\Lambda_b^0(udb)$.  Among these, the $\Lambda_c^+$ is the simplest hadron containing an up-type (charge $+2/3$) heavy quark~\cite{PDG}. Its free beta decay, with a final state including both a lighter hadron(s) and an antilepton-neutrino pair, is referred to as a semileptonic decay. This decay offers a clear view of the dynamics of the strong and weak interactions. 
The hadronic part can be well separated from the leptonic part and factorized with transition form-factors that encapsulate the dynamics of strong interactions, which provides robust validation for quantum chromodynamics (QCD) calculations.
In contrast, the leptonic part allows precisely determination the Cabibbo-Kobayashi-Maskawa (CKM) matrix~\cite{Cabibbo:1963yz,Kobayashi:1973fv} element $|V_{cd}|$, independently constraining this fundamental parameter of the weak interaction theory. 
\vspace{-1em}
\begin{figure}[h]
    \centering
    \includegraphics[width=0.45\textwidth]{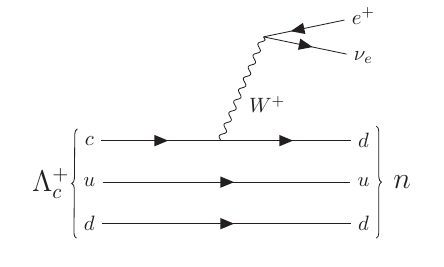}
    \caption{The leading-order Feynman diagram for $\beta^+$ decay of the charmed heavy baryon $\lambdacp$ into a neutron ($n$), positron ($e^+$), and electron neutrino ($\nu_e$) via an intermediate $W^+$ boson.}
    \label{fig:feynman}
\end{figure}

The experimental identification of the lightest charmed baryon, $\Lambda^+_c$, was accomplished more than 40 years ago~\cite{Knapp:1976qw,Abrams:1979iu}. Experimental studies of its semileptonic decays can be used to test various QCD-derived phenomenology models~\cite{NSR2021}. However, our understanding of its characteristics was initially quite limited. The situation began to change in 2014 when the BESIII experiment conducted the first measurement of the absolute branching fractions (BFs) of the $\Lambda^+_c$ decays~\cite{NSR2021,Ke:2023qzc,BESIII:2015ysy,BESIII:2015bjk} based on pair production of $\Lambda^+_c\Bar{\Lambda}^-_c$ just above the production energy threshold. Since then, the two Cabibbo-favored semileptonic decays, $\Lambda_c^+\to\Lambda l^{+}\nu_{l}$ ($l=e$, $\mu$), have been studied and their BFs are precisely measured, as well as the hadronic transition form factors~\cite{BESIII:2015ysy,BESIII:2016ffj,BESIII:2022ysa,BESIII:2023jxv} which describe the strong interaction effects in the decays. In contrast, another semileptonic decay, involving a neutron, $\Lambda_c^+\to n l^{+}\nu_{l}$, representing a Cabibbo-suppressed transition ($c\to W^+d$) as shown in Figure~\ref{fig:feynman}, has not been yet observed.  It is essentially certain to exist, and detailed calculations have been made based on Lattice QCD (LQCD) and massive QCD-derived phenomenology models~\cite{Meinel:2016dqj,Perez-Marcial:1989sch,Ivanov:1996fj,Pervin:2005ve,Gutsche:2014zna,Faustov:2016yza,Lu:2016ogy,Li:2016qai,Meinel:2017ggx,Geng:2017mxn,Zhao:2018zcb,Geng:2019bfz,Geng:2020fng,Geng:2020gjh,He:2021qnc,Geng:2022fsr,Zhang:2023nxl}. To test these predictions in different models, experimental results on the decay $\Lambda_c^+\to n e^+ \nu_e$ are desired. In addition, by combining results on the decay dynamics with the predicted hadronic transition form factor, the CKM matrix element $\left|V_{cd}\right|$ can be determined for the first time from charmed baryon.

\begin{figure}[h]%
    \centering
    \includegraphics[width=0.8\textwidth]{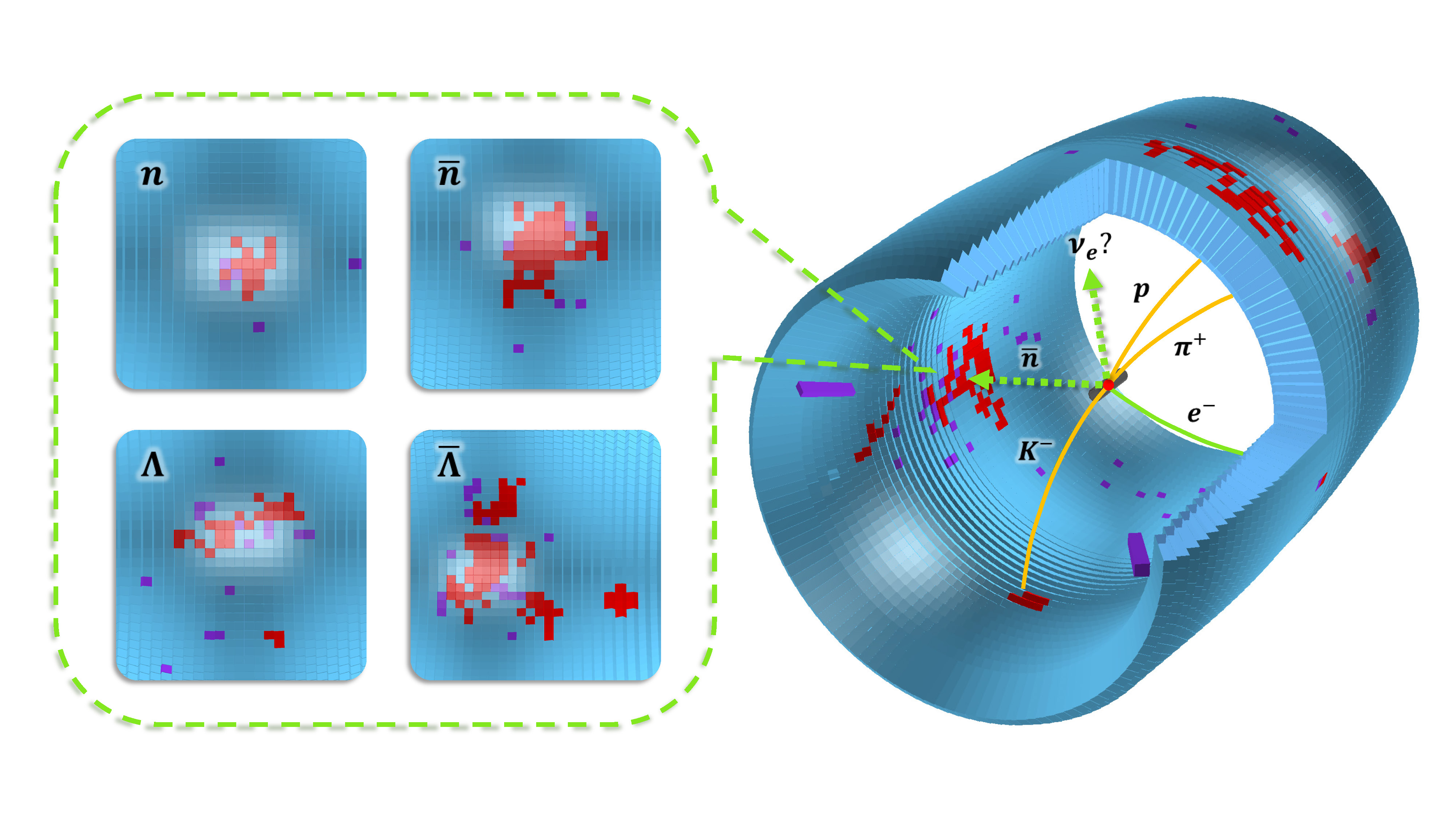}
    \caption{\textbf{(Right) Visualization of a $\Lambda^+_c\to pK^-\pi^+$, $\Bar{\Lambda}^-_c\to\bar{n}e^-\Bar{\nu}_e$ event in the BESIII detector~\cite{Huang:2022wuo,Li:2024pox}.} The blue cylindrical represents the barrel EMC crystal configuration, and the red and violet pixels mark the on-fire crystals. The EMC showers are clusters of adjacent active crystals defined by the BESIII EMC reconstruction algorithm~\cite{He:2011zzd}, with red pixels representing crystals within EMC showers and violet pixels being outside of them. (Left) The enclosed area displays zoomed-in views of the four typical EMC hit patterns from $n$, $\bar{n}$, $\Lambda(\to n \pi^0)$ and $\bar{\Lambda}(\to \bar{n} \pi^0)$, respectively.}
    \label{fig:vislmdenu}
\end{figure}

In practice, identifying the decay $\Lambda_c^{+}\to n e^{+}\nu_{e}$ faces great challenges at BESIII~\cite{BESIII:2009fln} or other similar particle physics experiments~\cite{BaBar:2001yhh,Belle:2000cnh,Belle-II:2018jsg,Achasov:2023gey}, because the neutral final state particles of the neutron and neutrino are hard to detect instrumentally. These particles cannot be reconstructed at all in BESIII's multilayer drift chamber, designed for charged particle tracking. Moreover, the ability to separate the signal process is strongly undermined due to the background process $\Lambda_c^{+}\to\Lambda e^{+}\nu_e$, whose BF is approximately ten times greater than that expected for $\Lambda_c^{+}\to n e^{+}\nu_e$. Here, the $\Lambda$ baryon can decay subsequently into $n\pi^0$, and the $\pi^0$ further decays into two photons. The detector response for the $n\pi^0$ particles in the background decay is very similar with that of the single neutron in the signal process, except for subtle differences in the pattern of deposited energy on the CsI(Tl) crystals of the electromagnetic calorimeter (EMC). Two extra photon showers are introduced in the $\Lambda\to n \pi^0$ background via the $\pi^0\to\gamma\gamma$ decays. The neutron showers, however, are more broadly dispersed than the photon showers and this often leads to the neutron showers blending in with photon showers or being mistaken for electronic noise, rendering the signal neutron indiscernible from the $\Lambda$ background. Figure~\ref{fig:vislmdenu} illustrates the shower patterns in the EMC for typical $\Lambda_c^{+}\to n e^{+}\nu_e$ signal events and $\Lambda^+_c\to\Lambda(\to n\pi^0)e^+\nu_e$ background events. Consequently, identifying signal events utilizing such patterns is almost impossible for common data analysis techniques in particle physics, even with most multivariate analysis tools~\cite{Hocker:2007ht}, such as boosted decision trees, not to mention less powerful traditional selection-based methods.

In this work, we report the first observation of the semileptonic decay $\Lambda_c^+\to n e^+\nu_{e}$ using $e^+e^-$ collision data collected with the BESIII detector, and the first measurement of the CKM matrix element $|V_{cd}|$ via a charmed baryon decay. To overcome the difficulties of signal identification and reconstruction, we resort to modern machine learning techniques like deep neural networks~\cite{deeplearning}, which have exhibited a powerful capability for learning relations and hidden patterns. A novel data-driven method is introduced for training and calibrating the deep neural network, utilizing the unprecedented sample of $10^{10}$ $J/\psi$ events at BESIII~\cite{BESIII:2021cxx}. This approach parallels recent advancements of jet tagging in LHC experiments~\cite{Larkoski:2017jix,Kogler:2018hem}, but at a new energy scale.

\section*{Results}
\subsection*{Candidates selection}
The BESIII experiment~\cite{BESIII:2009fln} is an electron-positron collider dedicated to study physics in the $\tau$-charm energy region~\cite{Asner:2008nq}, which is further described in the Methods. Data analyzed in this work consist of $e^+e^-$ collision data taken at seven center-of-mass energies between 4.600 $\mathrm{GeV}$ and 4.699 $\mathrm{GeV}$, corresponds to an integrated luminosity of 4.5~$\mathrm{fb}^{-1}$~\cite{BESIII:2022ulv}. At these energy points, $\Lambda^+_c$ and $\Bar{\Lambda}^-_c$ baryons are always produced in pairs without accompanying hadrons. This pristine production environment enables the utilization of a double-tag (DT) technique~\cite{MARK-III:1985hbd}, initially reconstructing either a $\bar{\Lambda}_c^-$ or $\Lambda_c^+$ baryon through its abundant hadronic decays, followed by the search for the signal decay in its recoiling partner. Consequently, the absolute signal BF can be accessed by
\begin{linenomath}
    \begin{equation}
        \mathcal{B}(\Lambda_c^+\to ne^+\nu_{e})=\frac{N_{\rm DT}}{N_{\rm ST}\cdot\epsilon_{\rm sig}},
        \label{eq:branch_main}
    \end{equation}
\end{linenomath}
where $N_{\rm ST}$ is the number of events finding the tagged $\bar{\Lambda}^-_c$ baryon, $N_{\rm DT}$ is the number of events finding both tagged $\Bar{\Lambda}^-_c$ and signal $\Lambda^+_c$ simultaneously, and $\epsilon_{\rm sig}$ is the corresponding signal detection efficiency. Throughout this Article, charge conjugation is implied by default unless explicitly stated. The detailed event selection criteria is described in the Methods, with the total number of tagged $\bar{\Lambda}^-_c$ baryons measured to be $N_{\rm ST}=105,506\pm399$.

\subsection*{Signal extraction via Graph Neural Network}
In processing the collision data with deep learning, we convert the deposited showers in EMC, not associated with any charged tracks or the $\bar{\Lambda}_c^-$ tag decay products, into a set of unordered nodes. Each node carries the measurable features of the shower, such as spatial coordinates, deposited energy, and the shower cluster profile. These nodes are organized as locally connected graphs to feed into a graph neural network (GNN) based on the ParticleNet architecture~\cite{Qu:2019gqs} which performs a binary classification between signal and background events.

A notable issue about the the (anti-)neutron is that its interaction with the EMC involve many complex mechanisms, such as annihilation, scattering, fusion, and capture, which are still poorly understood in the sub-GeV energy region~\cite{Pietropaolo:2020frm}. As a consequence, computer-based Monte-Carlo (MC) simulations of these interactions are unreliable. Owing to the unprecedented sample of $10^{10}$ $J/\psi$ events collected at BESIII~\cite{BESIII:2021cxx}, the real (anti-)neutron interactions in the EMC crystals can be calibrated in a data-driven approach, by selecting (anti-)neutron control samples of high purity and large statistics from the processes like $J/\psi\to\Bar{p}n\pi^+(p \Bar{n}\pi^-)$. In this work, we establish a data-driven procedure for training and calibrating the GNN model based on various neutron and $\Lambda$ control samples as follows. Note that, the two charge-conjugate channels are separately processed due to the very different interactions between neutrons and anti-neutrons with the detector material.

\begin{figure}[ht]%
    \centering
    \includegraphics[width=\textwidth]{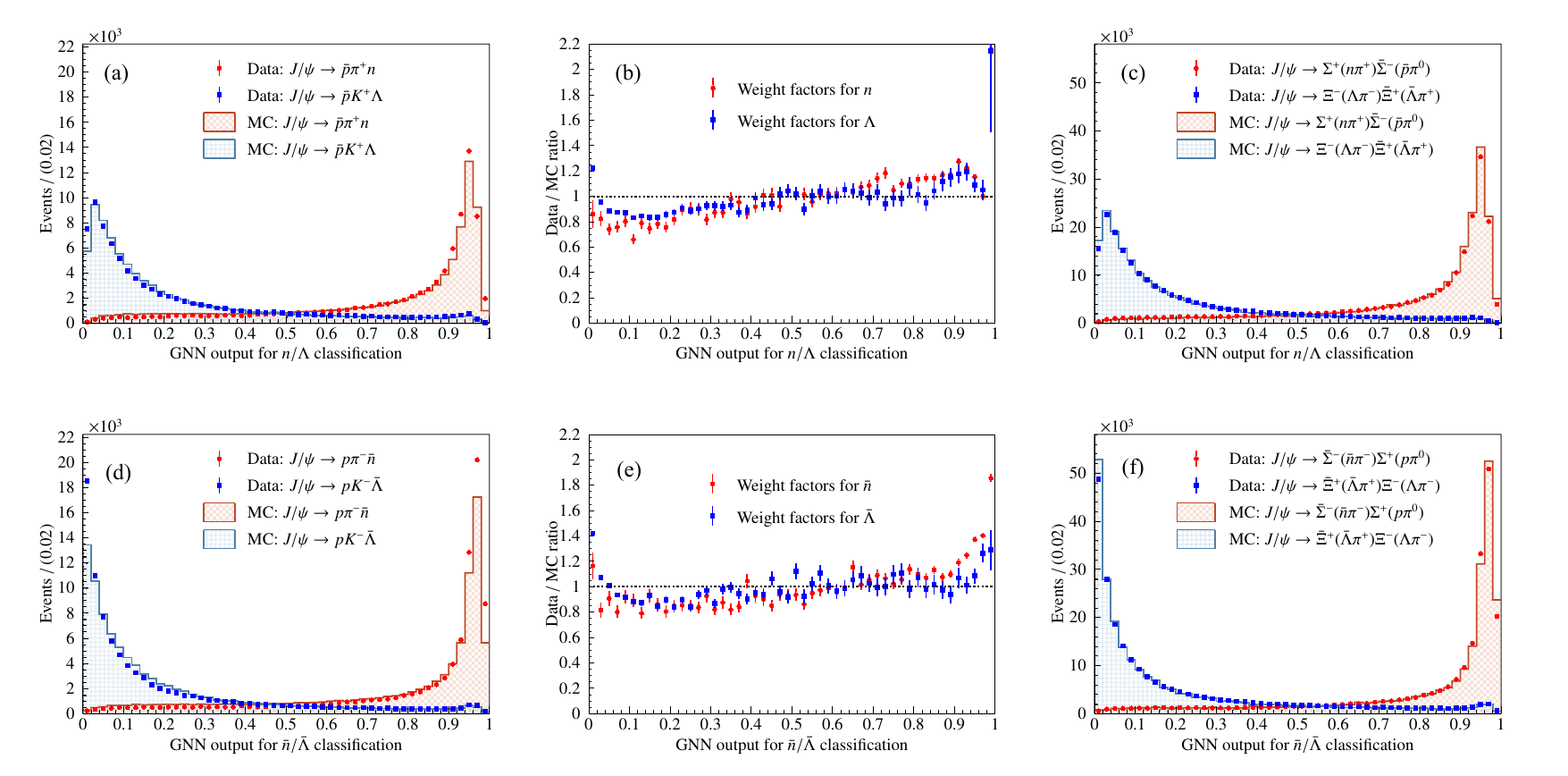}
    \caption{\textbf{The inference, calibration and validation of the GNN model.} (a, d) The GNN output distributions of $J/\psi\to\Bar{p}\pi^+ n$ and $J/\psi\to\Bar{p} K^+ \Lambda$ control samples prior to the MC corrections. (b, e) The normalized weight functions taken from the data-versus-MC ratios. (c, f) The GNN output distributions of $J/\psi\to\Sigma^+(\to n\pi^+) \, \Bar{\Sigma}^-(\to\Bar{p}\pi^0)$ and $J/\psi\to\Xi^+(\to \Lambda\pi^+) \, \Bar{\Xi}^-(\to\Bar{\Lambda}\pi^-)$ control samples post the MC corrections. Uncertainties on the data points are statistical only and represent one standard deviation.}\label{fig:train}
\end{figure}

\begin{figure}[h]%
    \centering
    \includegraphics[width=0.9\textwidth]{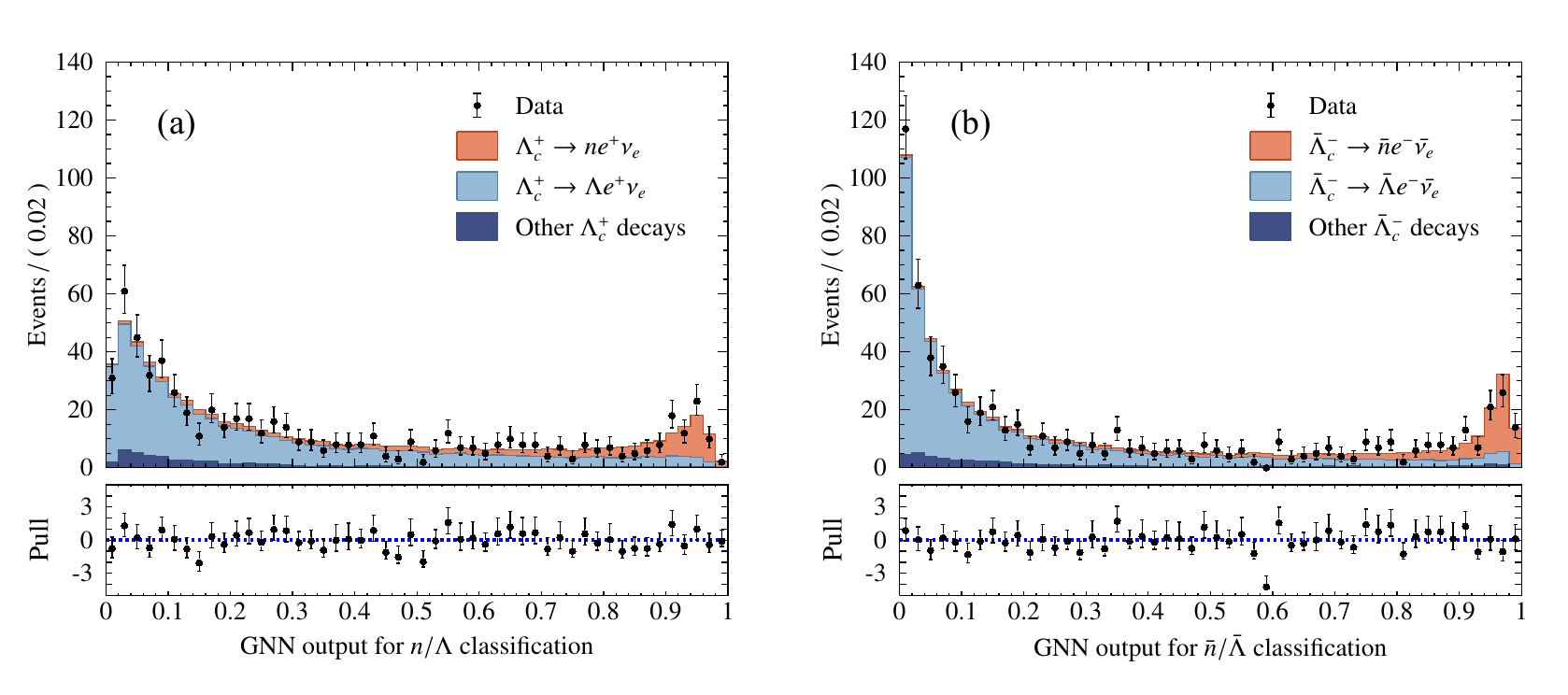}
    \caption{\textbf{The GNN output distributions in data.} (a) Fit to the GNN output distribution in $\bar{\Lambda}_c^{-}\to\bar{n} e^{-}\bar{\nu}_e$ signal candidates. (b) Fit to the GNN output distribution in $\Lambda_c^{+}\to n e^{+}\nu_e$ signal candidates. The error bars of data points are statistical only and represent one standard deviation. The stacked histograms show the total fitting results. The orange histograms represent the signal components, the light blue histograms represent the $\Lambda_c^{+}\to\Lambda e^{+}\nu_e$ or $\bar{\Lambda}_c^{-}\to\bar{\Lambda} e^{-}\bar{\nu}_e$ components, and the dark blue histograms represent other $\Lambda^+_c$ or $\bar{\Lambda}_c^{-}$ decay components.}\label{fig:fit}
\end{figure}

\begin{itemize}
    \item \textbf{Preparing the control samples.} We select neutron and $\Lambda$ control samples from $J/\psi\to\Bar{p}\pi^+ n$ and $J/\psi\to\Bar{p} K^+ \Lambda$ events, respectively, in BESIII real data at $J/\psi$ peak energy. After reconstructing the final-state $\bar{p}$ and $\pi^+$ or $K^+$, the control samples are purified by requiring the recoil mass $M_n$ ($M_{\Lambda}$) to be within the neutron ($\Lambda$) nominal mass region. The momentum range of the neutron ($\Lambda$) in the control samples covers that in the decay of $\Lambda_c^{+}\to n e^{+}\nu_{e}$ ($\Lambda_c^{+}\to\Lambda e^{+}\nu_e$). 
    The training sample for GNN is a random shuffle of the neutron and $\Lambda$ control samples with equal statistics, containing approximately 3.5 million events and with a purity greater than 99\%.

    \item \textbf{Organizing the data structure.} The identified physics-related showers deposited in EMC, not associated with the $\bar{p}$ and $\pi^+(K^+)$ in the neutron ($\Lambda$) control sample, are used to form the point cloud. Each point in the cloud carries definite low-level features of the shower, including azimuth angle in the laboratory frame, energy deposit in the EMC crystals, the number of crystals with energy above a minimum threshold, timing information, the ratio of deposited energy between the $3\times3$ and $5\times5$ crystal regions around the center (most energetic crystal) of the shower, the lateral and secondary moments as well as A20 and A42 Zernike moments~\cite{Sinkus:1996ch}. 
    
    \item \textbf{Building up the GNN model.} The architecture of the GNN model largely follows the original configurations in the ParticleNet~\cite{Qu:2019gqs}, consisting of three EdgeConv blocks~\cite{10.1145/3326362}, a global average pooling layer, and two fully connected layers. The number of nearest neighbors for all three EdgeConv blocks is set to 6, with varying numbers of channels, specifically $(8, 8, 8)$, $(16, 16, 16)$, and $(32, 32, 32)$, respectively. A channelwise global average pooling operation is applied after the EdgeConv blocks to aggregate the learned features over all points in the cloud, and then followed by a fully connected layer with 32 units and the ReLU activation. To prevent over-fitting, a dropout layer~\cite{JMLR:v15:srivastava14a} with a drop probability of 0.1 is included. A fully connected layer with two units, followed by a softmax function, is used to generate the output for a binary classification task.
    
    \item \textbf{Training the GNN model.} Training and optimization of the GNN model are performed using the open-source framework Weaver~\cite{weaver}, implemented with PyTorch~\cite{Paszke:2019xhz}. Events from the two sets of $J/\psi$ control samples are randomly selected with equal probability and mixed.  Then, 90\% of these events are used for training and 10\% are used for independent evaluation. The model is trained for 50 epochs with a batch size of 4096. The Lookahead optimizer~\cite{zhang2019lookahead} with $k=6$ and $\alpha=0.5$ is employed to minimize the cross-entropy loss, with the inner optimizer being RAdam~\cite{xiong2020layer} with $\beta_1=0.95$, $\beta_2=0.999$, and $\epsilon=10^{-5}$. The initial learning rate is 0.004, which remains constant for the first 70\% of the epochs, and then decays exponentially to 1\% of the initial value at the end of the training. 
    
    \item \textbf{Inference and calibration of the GNN model.} The resultant trained GNN model is applied to the selected EMC showers in both the $J/\psi$ and $\Lambda^+_c$ candidate events, which predicts a signal probability between 0 and 1 for each event. As is indicated in Figure~\ref{fig:train}(a,~d), discrepancies arise in the GNN output distributions between real data and MC simulations due to the imperfect modeling of decay dynamics and detector response. To address this issue, we take the relative data-versus-MC ratios in the $J/\psi$ control samples as normalized weighting functions $\omega({\rm output})=\mathcal{PDF}^{\rm Data}({\rm output})/\mathcal{PDF}^{\rm MC}({\rm output})$, as shown in Figure~\ref{fig:train}(b,~e), where $\mathcal{PDF}^{\rm Data}({\rm output})$ and $\mathcal{PDF}^{\rm MC}({\rm output})$ represent the normalized probability density functions (PDFs) for the GNN output distributions in data and MC simulation of the $J/\psi$ control samples, respectively. The MC-determined distributions for the $\Lambda^+_c$ signal and background channels are then corrected according to the weight functions, which agree well with the data as seen in Figure~\ref{fig:fit}. The residual effect of the data-MC discrepancy is considered as a systematic uncertainty source, and is discussed in the following sections.
\end{itemize}

Based on the trained and calibrated GNN model, Figure~\ref{fig:fit} illustrates the output probability distributions of the surviving $\Lambda^+_c$ candidates. Clear enhancements in the high and low probability ranges are visible, which arise from $\Lambda_c^{+}\to n e^{+}\nu_{e}$ signal events and $\Lambda_c^{+}\to\Lambda e^{+}\nu_e$ backgrounds events, respectively. To count signal events, simultaneous binned maximum-likelihood fits to the GNN output distributions are performed separately to the data for $\Lambda_c^{+}\to n e^{+}\nu_{e}$ and $\bar{\Lambda}_c^{-}\to \bar{n} e^{-}\bar{\nu}_{e}$. Assuming $CP$ conservation, the BFs for the two charge-conjugate signal channels are set to be equal. The PDFs used in the fit for $\Lambda_c^{+}\to n e^{+}\nu_{e}$ and $\Lambda_c^{+}\to\Lambda e^{+}\nu_e$ are modeled with templates from MC simulation corrected with the neutron and $\Lambda$ control samples, respectively. In addition, there is a small component of other $\Lambda^+_c$ decay backgrounds, whose contributions are fixed according to MC simulation. The yields for the $\Lambda_c^{+}\to n e^{+}\nu_{e}$ and $\Lambda_c^{+}\to\Lambda e^{+}\nu_e$ components are free parameters. Finally, we obtain the yields in the tagged events of $\Lambda_c^{+}\to n e^{+}\nu_{e}$ and its conjugate channel to be $134\pm13$ and $131\pm12$, respectively. The corresponding signal efficiencies, $\epsilon_{\rm sig}$, in Eq.~\eqref{eq:branch_main} are determined with dedicated MC simulation, as discussed in the Methods, to be $(70.09\pm0.20)\%$ for $\Lambda_c^{+}\to n e^{+}\nu_{e}$ and $(70.39\pm0.20)\%$ for its conjugate channel, respectively. The signal BF is determined via Eq.~\eqref{eq:branch_main} to be $(0.357\pm0.034)\%$,  where the uncertainty is statistical only. The statistical significance for the signal is over 10 standard deviations, based on Wilks' theorem~\cite{Wilks:1938dza}, marking the first observation of the process $\Lambda_c^{+}\to ne^+\nu_{e}$. As a validation check on the analysis strategy, the BF for the background process $\Lambda_c^{+}\to\Lambda e^{+}\nu_e$ is calculated to be $(3.55\pm0.14)\%$ from the simultaneous fit, consistent with the previous measurement~\cite{BESIII:2022ysa}. 

\subsection*{Systematic uncertainties}
Several sources of systematic uncertainty have been investigated and the total contribution is 4.0\% of the central BF value, as detailed in the Methods. In particular, we study two issues related to the robustness and reliability of the machine learning model: domain shift and network uncertainty. Domain shift~\cite{quinonero2008dataset} describes the mismatch between training samples and evaluation samples. In this work, it refers to the potential difference of EMC shower profiles between $J/\psi$ and $\Lambda^+_c$ data sets, due to the kinematic phase space or other underlying dependence. This deviation could bias the correction to MC-derived GNN outputs using the $J/\psi$ control sample, and therefore the fit in Figure~\ref{fig:fit}. To evaluate this effect, we perform the calibration procedure on another set of neutron and $\Lambda$ control samples based on different $J/\psi$ processes $J/\psi\to\Sigma^+(\to n\pi^+)\Bar{\Sigma}^-(\to\Bar{p}\pi^0)$ and $J/\psi\to\Xi^+(\to \Lambda\pi^+)\Bar{\Xi}^-(\to\Bar{\Lambda}\pi^-)$. As illustrated in Figure~\ref{fig:train}(c,~f), the effects of residual data-MC discrepancies for these control samples are small, despite the shower distributions differing from the neutron training sample in the $J/\psi$ control sample and the $\Lambda^+_c$ data sets, indicating the validity of our calibration method to the GNN model. Network uncertainty describes the systematic effect on the choice of the trained GNN model, which is estimated via the ensemble method~\cite{sagi2018ensemble} by combining the predictions of multiple different networks at inference.

\section*{Discussion}
\begin{figure}[tp]%
    \centering
    \includegraphics[width=0.9\textwidth]{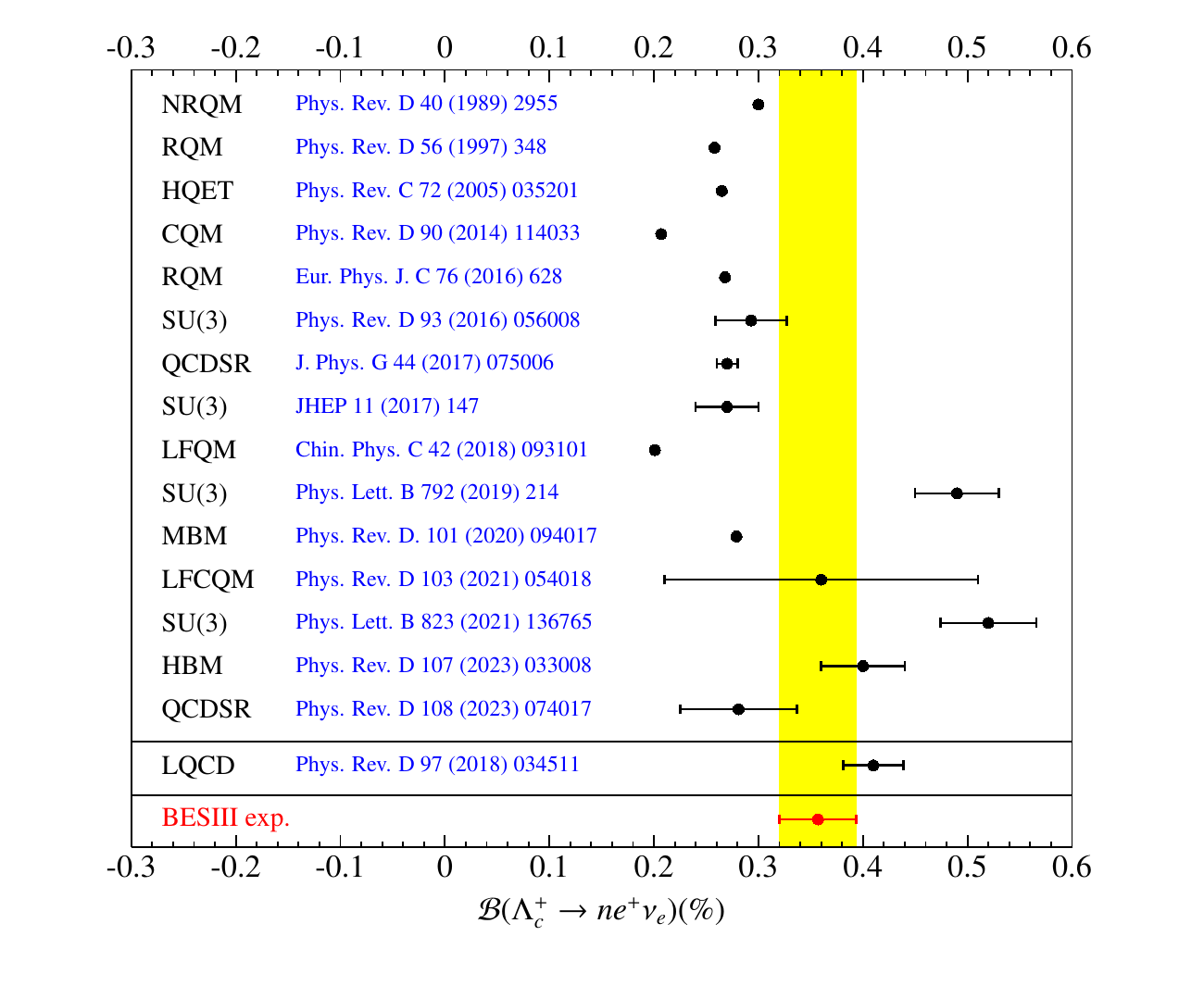}
    \caption{Comparison of our BF measurement with the theoretical predictions in Refs.~\cite{Perez-Marcial:1989sch,Ivanov:1996fj,Pervin:2005ve,Gutsche:2014zna,Faustov:2016yza,Lu:2016ogy,Li:2016qai,Meinel:2017ggx,Geng:2017mxn,Zhao:2018zcb,Geng:2019bfz,Geng:2020fng,Geng:2020gjh,He:2021qnc,Geng:2022fsr,Zhang:2023nxl}. The error bars represent one standard deviation of the BF results, calculated as a sum in quadrature of the statistical and systematic uncertainties. Note that some predictions do not report uncertainties.}\label{fig:comp}
\end{figure}

In conclusion, we report the first observation of a Cabibbo-suppressed $\Lambda_c^+$ beta decay into a neutron, $\Lambda_c^+ \rightarrow n e^+ \nu_{e}$, with a statistical significance of more than 10$\sigma$, based on $4.5~\mathrm{fb}^{-1}$ of electron-positron annihilation data collected with the BESIII detector in the energy region just above the $\Lambda^+_c\Bar{\Lambda}^-_c$ threshold. The machine learning technique employed exhibits a great capability for extracting small signals intermingled with very large and similarly-behaved backgrounds in experimental high energy physics; such a task is almost impossible with traditional selection-based methods. 
Meanwhile, we develop a validation pipeline to quantify and reduce systematic uncertainties associated with the machine learning model, leveraging abundant $J/\psi$ control samples collected at BESIII. The absolute branching fraction for the semileptonic decay $\Lambda_c^{+}\to n e^{+}\nu_{e}$ is measured to be
\begin{linenomath}
    \begin{equation}
        \mathcal{B}(\Lambda_c^+\to ne^+\nu_{e})=(0.357\pm0.034_{\rm stat.}\pm0.014_{\rm syst.})\%,
        \label{eq:result1}
    \end{equation}
\end{linenomath}
where the first uncertainty is statistical and the second is systematic. Our result demonstrates a level of precision comparable to the LQCD prediction~\cite{Meinel:2017ggx}, and is consistent with it within one standard deviation. The comparisons with other theoretical calculations~\cite{Perez-Marcial:1989sch,Ivanov:1996fj,Pervin:2005ve,Gutsche:2014zna,Faustov:2016yza,Lu:2016ogy,Li:2016qai,Meinel:2017ggx,Geng:2017mxn,Zhao:2018zcb,Geng:2019bfz,Geng:2020fng,Geng:2020gjh,He:2021qnc,Geng:2022fsr,Zhang:2023nxl} are shown in Figure~\ref{fig:comp}. The absence of detectors capable of accurately assessing neutron energy and position restricted us to precisely measure the transition form factors, which is relevant to the momentum transfer $q^2=(p_{\Lambda^+_c}-p_n)^2$. Still, the measured absolute BF provides significant insights, shedding light on the di-quark structure within the $\Lambda_c^+$ core and the $\pi-N$ clouds~\cite{Thomas:1981vc} in the low $q^2$ regime.

In addition, we present a measurement of the CKM matrix element $\left|V_{cd}\right|$ using a novel decay mode.  A recent LQCD calculation~\cite{Meinel:2017ggx} gives the $q^2$-integrated partial width of $\Lambda_c^{+}\to n e^{+}\nu_{e}$ as $\Gamma(\Lambda_c^{+}\to n e^{+}\nu_{e})=|V_{cd}|^{2}(0.405\pm0.016\pm0.020)\ \rm ps^{-1}$, where the uncertainties include statistical and systematic ones propagated from the predicted form factors.  Using current $\Lambda_c^+$ lifetime $\tau_{\Lambda_c^+}=(0.2032\pm0.0012)\,\rm ps$~\cite{Belle-II:2022ggx}, we extract the magnitude of $\left|V_{cd}\right|$ as
\begin{eqnarray}
    \left|V_{cd}\right| =0.208\pm0.011_{\rm exp.}\pm0.007_{\rm LQCD}\pm0.001_{\tau_{\Lambda_c^+}},
\end{eqnarray}
at a precision of 6\% and consistent with the world average value $(0.221\pm0.004)$~\cite{PDG}, which is determined with the charmed meson (semi-)leptonic decays and neutrino scattering. Future improvements on our precision would rely on more statistics of $\Lambda^+_c$ data collected at BESIII, as well as improved theoretical calculations of the involved form factors.

\clearpage
\section*{Methods}
\subsection*{Experimental apparatus}
The BESIII detector~\cite{BESIII:2009fln} records symmetric $e^+e^-$ collisions provided by the BEPCII storage ring~\cite{Yu:IPAC2016-TUYA01} in the center-of-mass energy ($\sqrt{s}$) range from 2.0 to 4.95~GeV, with a peak luminosity of $1\times10^{33}$~cm$^{-2}$s$^{-1}$ achieved at $\sqrt{s} = 3.77\;\text{GeV}$. BESIII has collected large data samples in this energy region~\cite{BESIII:2020nme,EcmsMea}. The cylindrical core of the BESIII detector covers 93\% of the full solid angle and consists of a helium-based multilayer drift chamber~(MDC), a plastic scintillator time-of-flight system~(TOF), and a CsI(Tl) electromagnetic calorimeter~(EMC), which are all enclosed in a superconducting solenoid magnet providing a 1.0~T magnetic field. The solenoid is supported by an octagonal flux-return yoke with resistive plate counter muon identification modules interleaved with steel. The charged-particle momentum resolution at $1~{\rm GeV}/c$ is $0.5\%$, and the ${\rm d}E/{\rm d}x$ resolution is $6\%$ for electrons from Bhabha scattering. The EMC measures photon energies with a resolution of $2.5\%$ ($5\%$) at $1$~GeV in the barrel (end cap) region. The time resolution in the TOF barrel region is 68~ps, while that in the end cap region is 110~ps. The end cap TOF system was upgraded in 2015 using multi-gap resistive plate chamber technology, providing a time resolution of 60~ps~\cite{etof1,etof2,etof3}; about 87\% of the data used here benefits from this upgrade. 

\subsection*{Monte Carlo simulation}
Simulated MC samples produced with a {\sc geant4}-based~\cite{GEANT4:2002zbu} package, which includes the geometric description of the BESIII detector and the detector response, are used to determine detection efficiencies and to estimate backgrounds. The simulation models the beam energy spread and initial state radiation (ISR) in the $e^+e^-$ annihilations with the generator {\sc kkmc}~\cite{Jadach:2000ir,Jadach:1999vf}. The inclusive MC sample includes the production of open charm processes, the ISR production of vector charmonium(-like) states, and the continuum processes incorporated in {\sc kkmc}. All particle decays are modelled with {\sc evtgen}~\cite{Lange:2001uf,Ping:2008zz} using BFs either taken from the Particle Data Group~\cite{PDG}, when available, or otherwise estimated with {\sc lundcharm}~\cite{Chen:2000tv,Yang:2014vra}. Final state radiation~(FSR) from charged final state particles is incorporated using the {\sc photos} package~\cite{Richter-Was:1992hxq}. The simulations of the decay $\Lambda_c^+\rightarrow n e^+\nu_e$ and $\Lambda_c^+\rightarrow \Lambda e^+\nu_e$ take into account their form factors, as predicted by LQCD~\cite{Meinel:2017ggx,Meinel:2016dqj}.

\subsection*{Event selection criteria}
The DT analysis approach allows for a straightforward and clean measurement of signal BF without knowledge of the total number of $\Lambda^+_c\Bar{\Lambda}^-_c$ events produced. The $\bar{\Lambda}_c^-$ baryon is firstly reconstructed in the ten exclusive hadronic decay modes $\bar{\Lambda}^-_c\rightarrow \bar{p} K^0_S$, $\bar{p} K^+\pi^-$, $\bar{p}K^0_S\pi^0$, $\bar{p} K^0_S\pi^+\pi^-$, $\bar{\Lambda}\pi^-$, $\bar{\Lambda}\pi^-\pi^0$, $\bar{\Lambda}\pi^-\pi^+\pi^-$, $\bar{\Sigma}^0\pi^-$, $\bar{\Sigma}^-\pi^0$, and $\bar{\Sigma}^-\pi^+\pi^-$. The intermediate particles $K^0_S$, $\pi^0$, $\bar{\Lambda}$, $\bar{\Sigma}^0$ and $\bar{\Sigma}^-$ are reconstructed via their dominant decay modes $K^0_S\rightarrow \pi^+\pi^-$, $\bar{\Lambda}\rightarrow\bar{p}\pi^+$, $\bar{\Sigma}^0\rightarrow \gamma\bar{\Lambda}$ with $\bar{\Lambda}\rightarrow \bar{p}\pi^+$, $\bar{\Sigma}^-\rightarrow \bar{p}\pi^0$, and $\pi^0\rightarrow \gamma\gamma$. The details of $\bar{\Lambda}_c^-$ reconstruction follow the method in Ref.~\cite{BESIII:2022xne}, and the selected sample is referred to as the single-tag (ST) sample. The signal decay $\Lambda_c^+\to ne^+\nu_{e}$ is then searched for in the system recoiling against the ST $\bar{\Lambda}_c^-$ baryon; successful tag plus signal candidates are referred to as DT events.

The signal BF is determined with Eq.~\eqref{eq:branch_main}. Here, $N_{\rm DT}$ is the yield of DT events. $N_{\rm ST}=\sum_{i,j}N^{i,j}_{\rm ST}$ is the total yield of ST $\bar{\Lambda}^-_c$ baryons, summing over the ST yields $N^{i,j}_{\rm ST}$ in the $i^{\rm th}$ ST mode at the $j^{\rm th}$ energy point.
The effective signal efficiency, $\epsilon_{\rm sig} = \sum_{i,j}(N^{i,j}_{\rm ST}\epsilon^{i,j}_{\rm DT}/\epsilon^{i,j}_{\rm ST})/N_{\rm ST}$, for selecting the signal decay in the presence of an ST $\bar{\Lambda}_c^-$ baryon, is averaged over the different ST modes and energy points. Here, $\epsilon^{i,j}_{\rm ST}$ and $\epsilon^{i,j}_{\rm DT}$ are the detection efficiencies of the ST $\bar{\Lambda}_c^-$ baryons and the DT candidates in the $i^{\rm th}$ ST mode at the $j^{\rm th}$ energy point, respectively. The results of ST yields $N^{i,j}_{\rm ST}$ are obtained following Ref.~\cite{BESIII:2022xne}, and are given in Table~\ref{tab:styield}. The ST and DT efficiencies, estimated with MC simulation, are listed in Tables~\ref{tab:steff} and \ref{tab:dteff}, respectively.

\begin{table}[htbp]
    \scriptsize
    \centering
    \caption{ST yields $N^{i,j}_{\rm ST}$ in the $i^{\rm th}$ ST mode at the $j^{\rm th}$ energy point.}\label{tab:styield}
    \begin{tabular}{l|c|c|c|c|c|c|c}
        \toprule
        modes $\lambdacp\to$ & $4.600\gev$ & $4.612\gev$ & $4.628\gev$ & $4.641\gev$ & $4.661\gev$ & $4.682\gev$  & $4.699\gev$ \\
        \midrule
        $\modea$             & $645\pm26$  & $113\pm11$  & $515\pm24$  & $557\pm25$  & $537\pm24$  & $1645\pm43$  & $458\pm23$  \\
        $\modeb$             & $3295\pm65$ & $592\pm28$  & $2909\pm62$ & $3136\pm64$ & $3025\pm62$ & $8572\pm104$ & $2486\pm56$ \\
        $\modec$             & $291\pm24$  & $65\pm12$   & $300\pm28$  & $288\pm26$  & $290\pm27$  & $870\pm46$   & $224\pm25$  \\
        $\moded$             & $321\pm26$  & $46\pm11$   & $261\pm25$  & $252\pm24$  & $297\pm25$  & $760\pm42$   & $232\pm24$  \\
        $\modeaa$            & $377\pm20$  & $64\pm8$    & $330\pm20$  & $360\pm20$  & $327\pm19$  & $1049\pm34$  & $259\pm17$  \\
        $\modebb$            & $858\pm40$  & $146\pm16$  & $750\pm37$  & $823\pm40$  & $727\pm36$  & $2204\pm63$  & $636\pm35$  \\
        $\modedd$            & $418\pm27$  & $80\pm12$   & $297\pm24$  & $375\pm26$  & $428\pm33$  & $1040\pm45$  & $321\pm25$  \\
        $\modeaaa$           & $250\pm18$  & $53\pm8$    & $171\pm15$  & $211\pm17$  & $223\pm17$  & $733\pm30$   & $175\pm15$  \\
        $\modeccc$           & $167\pm18$  & $43\pm11$   & $149\pm17$  & $152\pm18$  & $131\pm17$  & $456\pm32$   & $120\pm17$  \\
        $\modeddd$           & $587\pm34$  & $125\pm17$  & $438\pm32$  & $560\pm36$  & $495\pm34$  & $1515\pm62$  & $479\pm37$  \\
        \bottomrule
    \end{tabular}
    \begin{tabular}{l|c|c|c|c|c|c|c}
        \toprule
        modes $\lambdacm\to$ & $4.600\gev$ & $4.612\gev$ & $4.628\gev$ & $4.641\gev$ & $4.661\gev$ & $4.682\gev$  & $4.699\gev$ \\
        \midrule
        $\Modea$             & $633\pm26$  & $126\pm12$  & $540\pm25$  & $552\pm25$  & $582\pm25$  & $1734\pm44$  & $501\pm24$  \\
        $\Modeb$             & $3516\pm64$ & $576\pm27$  & $2992\pm62$ & $3125\pm63$ & $2924\pm60$ & $8970\pm104$ & $2699\pm57$ \\
        $\Modec$             & $318\pm24$  & $62\pm11$   & $296\pm24$  & $315\pm25$  & $298\pm24$  & $922\pm43$   & $245\pm23$  \\
        $\Moded$             & $292\pm23$  & $60\pm11$   & $235\pm21$  & $276\pm22$  & $260\pm22$  & $788\pm38$   & $234\pm21$  \\
        $\Modeaa$            & $380\pm20$  & $56\pm8$    & $346\pm20$  & $345\pm20$  & $344\pm20$  & $1028\pm34$  & $280\pm18$  \\
        $\Modebb$            & $888\pm39$  & $164\pm17$  & $730\pm36$  & $798\pm37$  & $770\pm36$  & $2202\pm61$  & $685\pm34$  \\
        $\Modedd$            & $355\pm24$  & $58\pm10$   & $291\pm22$  & $374\pm25$  & $349\pm24$  & $1048\pm42$  & $330\pm24$  \\
        $\Modeaaa$           & $276\pm19$  & $49\pm8$    & $243\pm16$  & $237\pm18$  & $233\pm18$  & $670\pm29$   & $197\pm16$  \\
        $\Modeccc$           & $149\pm17$  & $31\pm7$    & $119\pm16$  & $143\pm17$  & $168\pm18$  & $432\pm30$   & $132\pm17$  \\
        $\Modeddd$           & $621\pm39$  & $95\pm15$   & $561\pm33$  & $520\pm34$  & $558\pm34$  & $1616\pm60$  & $480\pm33$  \\
        \bottomrule
    \end{tabular}
\end{table}

\begin{table}[tp]
    \scriptsize
    \centering
    \caption{ST detection efficiencies $\epsilon^{i,j}_{\rm ST} (\%)$ in the $i^{\rm th}$ ST mode at the $j^{\rm th}$ energy point.}\label{tab:steff}
    \begin{tabular}{l|c|c|c|c|c|c|c}
        \toprule
        modes $\lambdacp\to$ & $4.600\gev$  & $4.612\gev$  & $4.628\gev$  & $4.641\gev$  & $4.661\gev$  & $4.682\gev$  & $4.699\gev$  \\
        \midrule
        $\modea$             & $56.1\pm0.3$ & $53.4\pm0.8$ & $51.8\pm0.3$ & $50.7\pm0.3$ & $49.7\pm0.3$ & $48.6\pm0.2$ & $47.6\pm0.3$ \\
        $\modeb$             & $51.5\pm0.1$ & $51.2\pm0.3$ & $49.4\pm0.1$ & $49.1\pm0.1$ & $48.4\pm0.1$ & $47.5\pm0.1$ & $47.0\pm0.1$ \\
        $\modec$             & $22.7\pm0.2$ & $23.0\pm0.6$ & $20.9\pm0.2$ & $20.8\pm0.2$ & $19.7\pm0.2$ & $19.2\pm0.1$ & $18.6\pm0.2$ \\
        $\moded$             & $24.0\pm0.3$ & $21.5\pm0.6$ & $21.5\pm0.3$ & $21.8\pm0.3$ & $21.4\pm0.3$ & $22.0\pm0.2$ & $19.4\pm0.3$ \\
        $\modeaa$            & $47.6\pm0.4$ & $45.5\pm0.9$ & $41.6\pm0.4$ & $40.5\pm0.4$ & $40.1\pm0.4$ & $40.1\pm0.2$ & $37.9\pm0.4$ \\
        $\modebb$            & $20.8\pm0.1$ & $18.9\pm0.3$ & $18.5\pm0.1$ & $18.6\pm0.1$ & $18.4\pm0.1$ & $17.6\pm0.1$ & $17.5\pm0.1$ \\
        $\modedd$            & $16.0\pm0.2$ & $13.7\pm0.4$ & $14.1\pm0.2$ & $14.4\pm0.2$ & $14.2\pm0.2$ & $14.2\pm0.1$ & $14.8\pm0.2$ \\
        $\modeaaa$           & $28.0\pm0.3$ & $24.5\pm0.8$ & $25.8\pm0.3$ & $25.2\pm0.3$ & $25.4\pm0.3$ & $24.7\pm0.2$ & $23.4\pm0.3$ \\
        $\modeccc$           & $22.8\pm0.4$ & $21.5\pm0.8$ & $22.4\pm0.4$ & $24.9\pm0.4$ & $22.4\pm0.4$ & $22.2\pm0.2$ & $21.4\pm0.4$ \\
        $\modeddd$           & $25.1\pm0.2$ & $25.2\pm0.5$ & $23.2\pm0.2$ & $22.8\pm0.2$ & $22.9\pm0.2$ & $22.3\pm0.1$ & $22.1\pm0.2$ \\
        \bottomrule
    \end{tabular}
    \begin{tabular}{l|c|c|c|c|c|c|c}
        \toprule
        modes $\lambdacm\to$ & $4.600\gev$  & $4.612\gev$  & $4.628\gev$  & $4.641\gev$  & $4.661\gev$  & $4.682\gev$  & $4.699\gev$  \\
        \midrule
        $\Modea$             & $56.3\pm0.3$ & $54.0\pm0.8$ & $51.8\pm0.3$ & $50.9\pm0.3$ & $49.6\pm0.3$ & $48.7\pm0.2$ & $47.6\pm0.3$ \\
        $\Modeb$             & $51.4\pm0.1$ & $51.0\pm0.3$ & $49.2\pm0.1$ & $48.2\pm0.1$ & $48.2\pm0.1$ & $46.8\pm0.1$ & $45.7\pm0.1$ \\
        $\Modec$             & $23.3\pm0.2$ & $21.6\pm0.6$ & $20.8\pm0.2$ & $20.9\pm0.2$ & $20.7\pm0.2$ & $20.4\pm0.1$ & $19.5\pm0.2$ \\
        $\Moded$             & $23.1\pm0.3$ & $22.2\pm0.6$ & $19.9\pm0.3$ & $20.1\pm0.3$ & $20.8\pm0.3$ & $19.6\pm0.2$ & $19.8\pm0.3$ \\
        $\Modeaa$            & $49.2\pm0.4$ & $48.4\pm0.9$ & $44.6\pm0.4$ & $45.2\pm0.4$ & $43.3\pm0.4$ & $42.6\pm0.2$ & $40.8\pm0.4$ \\
        $\Modebb$            & $21.8\pm0.1$ & $20.7\pm0.3$ & $19.8\pm0.1$ & $19.6\pm0.1$ & $19.4\pm0.1$ & $18.8\pm0.1$ & $18.5\pm0.1$ \\
        $\Modedd$            & $15.3\pm0.2$ & $13.5\pm0.4$ & $13.7\pm0.2$ & $14.1\pm0.2$ & $14.0\pm0.2$ & $13.8\pm0.1$ & $14.5\pm0.2$ \\
        $\Modeaaa$           & $30.9\pm0.4$ & $28.9\pm0.8$ & $28.7\pm0.4$ & $27.1\pm0.3$ & $27.6\pm0.4$ & $27.2\pm0.2$ & $25.3\pm0.4$ \\
        $\Modeccc$           & $24.5\pm0.4$ & $23.6\pm0.9$ & $24.5\pm0.4$ & $24.8\pm0.4$ & $24.0\pm0.4$ & $23.1\pm0.2$ & $23.0\pm0.4$ \\
        $\Modeddd$           & $25.8\pm0.2$ & $26.3\pm0.5$ & $23.7\pm0.2$ & $23.9\pm0.2$ & $23.4\pm0.2$ & $22.4\pm0.1$ & $22.9\pm0.2$ \\
        \bottomrule
    \end{tabular}
\end{table}

\begin{table}[htbp]
    \scriptsize
    \centering
    \caption{DT detection efficiencies $\epsilon^{i,j}_{\rm DT} (\%)$ in the $i^{\rm th}$ ST mode at the $j^{\rm th}$ energy point.}\label{tab:dteff}
    \begin{tabular}{l|c|c|c|c|c|c|c}
        \toprule
        modes $\lambdacp\to$ & $4.600\gev$    & $4.612\gev$    & $4.628\gev$    & $4.641\gev$    & $4.661\gev$    & $4.682\gev$    & $4.699\gev$    \\
        \midrule
        $\modea$             & $39.39\pm0.34$ & $37.43\pm0.34$ & $36.07\pm0.33$ & $36.07\pm0.33$ & $34.81\pm0.33$ & $34.87\pm0.33$ & $33.40\pm0.33$ \\
        $\modeb$             & $35.53\pm0.33$ & $34.25\pm0.33$ & $34.20\pm0.33$ & $33.55\pm0.33$ & $34.07\pm0.33$ & $32.56\pm0.32$ & $31.67\pm0.32$ \\
        $\modec$             & $16.29\pm0.18$ & $16.42\pm0.18$ & $15.42\pm0.18$ & $15.54\pm0.18$ & $15.13\pm0.17$ & $14.89\pm0.17$ & $14.61\pm0.17$ \\
        $\moded$             & $16.17\pm0.18$ & $15.15\pm0.18$ & $14.72\pm0.17$ & $14.47\pm0.17$ & $14.59\pm0.17$ & $14.27\pm0.17$ & $14.05\pm0.17$ \\
        $\modeaa$            & $33.74\pm0.33$ & $31.94\pm0.32$ & $30.20\pm0.32$ & $29.94\pm0.32$ & $29.28\pm0.31$ & $27.84\pm0.31$ & $27.29\pm0.31$ \\
        $\modebb$            & $15.03\pm0.17$ & $14.11\pm0.17$ & $13.74\pm0.17$ & $13.45\pm0.17$ & $13.05\pm0.17$ & $12.96\pm0.16$ & $12.79\pm0.17$ \\
        $\modedd$            & $10.51\pm0.11$ & $9.82\pm0.10$  & $9.62\pm0.10$  & $9.63\pm0.10$  & $9.84\pm0.10$  & $9.68\pm0.10$  & $9.52\pm0.10$  \\
        $\modeaaa$           & $20.87\pm0.20$ & $19.66\pm0.20$ & $18.41\pm0.19$ & $18.43\pm0.19$ & $17.81\pm0.19$ & $17.67\pm0.19$ & $16.92\pm0.19$ \\
        $\modeccc$           & $17.78\pm0.15$ & $17.96\pm0.15$ & $17.24\pm0.15$ & $17.13\pm0.15$ & $16.71\pm0.15$ & $16.13\pm0.15$ & $15.83\pm0.15$ \\
        $\modeddd$           & $18.17\pm0.15$ & $18.21\pm0.15$ & $17.49\pm0.15$ & $17.04\pm0.15$ & $16.96\pm0.15$ & $16.45\pm0.15$ & $16.10\pm0.15$ \\
        \bottomrule
    \end{tabular}
    \begin{tabular}{l|c|c|c|c|c|c|c}
        \toprule
        modes $\lambdacm\to$ & $4.600\gev$    & $4.612\gev$    & $4.628\gev$    & $4.641\gev$    & $4.661\gev$    & $4.682\gev$    & $4.699\gev$    \\
        \midrule
        $\Modea$             & $42.38\pm0.34$ & $40.23\pm0.34$ & $39.69\pm0.34$ & $39.21\pm0.34$ & $37.70\pm0.34$ & $37.11\pm0.33$ & $35.69\pm0.34$ \\
        $\Modeb$             & $33.81\pm0.33$ & $34.66\pm0.33$ & $33.37\pm0.33$ & $33.47\pm0.33$ & $32.78\pm0.32$ & $32.01\pm0.32$ & $31.49\pm0.32$ \\
        $\Modec$             & $16.36\pm0.18$ & $16.03\pm0.18$ & $15.65\pm0.18$ & $15.62\pm0.18$ & $15.47\pm0.18$ & $15.08\pm0.18$ & $14.85\pm0.18$ \\
        $\Moded$             & $14.06\pm0.17$ & $12.88\pm0.16$ & $12.99\pm0.17$ & $12.84\pm0.16$ & $13.01\pm0.16$ & $13.02\pm0.17$ & $12.79\pm0.16$ \\
        $\Modeaa$            & $35.01\pm0.33$ & $34.85\pm0.33$ & $33.89\pm0.33$ & $32.71\pm0.32$ & $32.09\pm0.32$ & $31.48\pm0.32$ & $30.62\pm0.32$ \\
        $\Modebb$            & $14.69\pm0.17$ & $14.81\pm0.18$ & $14.26\pm0.17$ & $13.71\pm0.17$ & $13.40\pm0.17$ & $13.79\pm0.17$ & $13.19\pm0.17$ \\
        $\Modedd$            & $8.97\pm0.10$  & $8.81\pm0.10$  & $8.54\pm0.10$  & $8.45\pm0.10$  & $8.79\pm0.10$  & $8.71\pm0.10$  & $8.72\pm0.10$  \\
        $\Modeaaa$           & $22.38\pm0.20$ & $21.88\pm0.20$ & $20.77\pm0.20$ & $20.93\pm0.20$ & $20.31\pm0.20$ & $19.63\pm0.20$ & $19.00\pm0.19$ \\
        $\Modeccc$           & $20.11\pm0.16$ & $20.08\pm0.16$ & $19.19\pm0.16$ & $18.53\pm0.16$ & $18.50\pm0.15$ & $18.07\pm0.15$ & $17.36\pm0.15$ \\
        $\Modeddd$           & $18.12\pm0.15$ & $18.21\pm0.15$ & $17.73\pm0.15$ & $17.56\pm0.15$ & $17.13\pm0.15$ & $16.44\pm0.15$ & $16.23\pm0.15$ \\
        \bottomrule
    \end{tabular}
\end{table}

DT candidates for $\Lambda^+_c\rightarrow ne^+\nu_e$ are selected by requiring exactly one remaining charged track, beyond the tag mode, with charge opposite to the tagged $\bar{\Lambda}^-_c$. The cosine of its emission angle ($\theta$) with respect to the beam direction is required within $\left|\cos\theta\right|<0.93$. The distance of the closest approach to the interaction point (IP) are required to be within $\pm$10~cm along the beam direction and 1~cm in the plane perpendicular to the beam. For particle identification, the information measured by MDC, TOF, and EMC are used to construct likelihoods for positron, pion and kaon hypotheses denoted as $\mathcal{L}(e)$, $\mathcal{L}(\pi)$ and $\mathcal{L}(K)$. The positron candidate must satisfy $\mathcal{L}(e)>0.001$ and $\mathcal{L}(e)/(\mathcal{L}(e)+\mathcal{L}(\pi)+\mathcal{L}(K))>0.8$. To further suppress the background, the ratio of the deposited energy in the EMC and the momentum from the MDC is required to be larger than 0.5.  

The remaining showers in the EMC, neither associated with any charged tracks nor used in the ST reconstruction, are analyzed further. To remove showers from electronic noise, the EMC shower time with respect to the event start time should be within $[0,\ 700]$ ns. At least one shower candidate is required as a candidate for the neutron from the signal decay. After the above selections, the dominant background component is found to be $\Lambda^+_c \rightarrow \Lambda e^+\nu_e$ with $\Lambda \rightarrow n\pi^0$. The contribution from non-$\Lambda^+_c\bar{\Lambda}_c^-$ hadronic background is negligible.

\subsection*{Systematic uncertainties}
The relevant sources of systematic uncertainties are summarized in Table~\ref{tab:syserr} and described as follows. Most systematic uncertainties related to ST selection cancel in the calculation of the signal BF, where the remaining uncertainty mainly comes from the uncertainty of the ST yields as 1.0\%~\cite{BESIII:2022xne}. The effect of a data-MC difference in the positron tracking efficiency is evaluated to be 0.3\% using the control sample $e^+e^-\to\gamma e^+e^-$ collected at $\sqrt{s}=3.097\gev$. Similarly, the effect of a data-MC difference in the positron identification efficiency is studied using the same $e^+e^-\to\gamma e^+e^-$ sample to be 1.2\%. Note these uncertainties are also applicable to the charge-conjugated electron. A data-MC efficiency difference from the ``no extra charged track" requirement is estimated using a control sample of DT $\Lambda^+_c\to nK^-\pi^+\pi^+$ collected at $\sqrt{s}=4.600\sim4.699\gev$, and is determined to be 1.1\%. Another data-MC efficiency difference due to the ``at least one shower candidate" requirement is calculated as 2.5\% using a control sample of DT $\Lambda^+_c\to nK^0_S\pi^+$ collected at $\sqrt{s}=4.600\sim4.699\gev$. For the MC model uncertainty, form factors provided by the LQCD~\cite{Meinel:2017ggx} are used to describe the dynamics of the signal process in determining the signal DT efficiency. Different MC model assumptions would alter the kinematic distributions of outgoing particles, and thus the signal efficiency when considering the detailed responses of BESIII detector. Other theoretical models~\cite{Perez-Marcial:1989sch,Faustov:2016yza,Gutsche:2014zna,Li:2016qai,Geng:2019bfz} are considered as variations and their corresponding signal efficiencies are calculated. Their standard deviation is taken as the systematic uncertainty to be 0.6\%. The binomial uncertainty in the signal efficiency due to finite size of signal MC sample, 0.2\%, is included as a systematic uncertainty.

To investigate the impact of domain shift in the simultaneous fit, control samples of $J/\psi\to\Sigma^+(\to n\pi^+) \, \Bar{\Sigma}^-(\to\Bar{p}\pi^0)$ and $J/\psi\to\Xi^+(\to \Lambda\pi^+) \, \Bar{\Xi}^-(\to\Bar{\Lambda}\pi^-)$ are selected from both real data and MC simulation. Figure~\ref{fig:train}(c,~e) compare the GNN output distributions for data and MC simulation after the correction procedure, which agree well with each other in large event statistics. A pseudo-data set is created by merging the two control samples with the yield ratio same as the ratio of the signals and backgrounds in the DT candidates in Figure~\ref{fig:fit}. The MC-determined shapes with corrections are adopted in fitting to the pseudo-data. To mitigate the effects of statistical fluctuations, a bootstrap re-sampling method~\cite{chernick2011bootstrap} is utilized. The output distribution of the fitted neutron yields is found to be consistent with the input yield within statistical uncertainty, and the deviation of the average value from the input value, 0.9\%, is taken as the systematic uncertainty due to the domain shift effect.

The GNN model uncertainty is quantified via the ensemble method, where a total of one hundred GNN models are trained independently. Among the different GNN settings, network weight initialization, batch processing sequence and dropout layer~\cite{JMLR:v15:srivastava14a} are randomly changed. The resultant signal BFs from the different trained GNN models follow a Gaussian distribution, where the BF with center value closest to the mean value of the Gaussian is chosen as the reported result. The difference between the chosen model and the Gaussian mean is negligible. The standard deviation of the Gaussian, 1.8\%, is taken as the systematic uncertainty.

The uncertainty related to the simultaneous yield fit is estimated by varying the details of the fitting procedure. The corrected MC-determined signal and background shapes are varied according to the relevant statistical fluctuations, due to the uncertainties of the correction function and the MC samples. The component of other $\Lambda^+_c$ decays is removed in an alternative fit.  The bootstrap re-sampling method mentioned above is again employed. The deviation of the mean value from the nominal fit is taken as the corresponding systematic uncertainty to be 1.2\%.

\begin{table}[h]
    \centering
    \small
    \caption{Summary of systematic uncertainties.}\label{tab:syserr}%
    \begin{tabular}{@{}lc@{}}
        \toprule
        Source                                & Relative uncertainty (\%) \\
        \midrule
        Single tag yields                     & 1.0                       \\
        Positron tracking                     & 0.3                       \\
        Positron identification               & 1.2                       \\
        No extra charged track requirement    & 1.1                       \\
        Neutron-induced shower reconstruction & 2.5                       \\
        MC model                              & 0.6                       \\
        MC statistics                         & 0.2                       \\
        Domain shift                          & 0.9                       \\
        GNN model                             & 1.8                       \\
        Simultaneous yield fit                & 1.2                       \\
        \midrule
        Total                                 & 4.0                       \\
        \bottomrule
    \end{tabular}
\end{table}

\section*{Data availability}
The raw data generated in this study have been deposited in the Institude of High Energy Physics mass storage silo database. The source data are available under restricted access for the complexity and large size, and the access can be obtained by contacting to besiii-publications@ihep.ac.cn. A minimum dataset to verify the result presented in the paper is available at {\sc zenodo} repository \href{https://doi.org/10.5281/zenodo.14048411}{https://doi.org/10.5281/zenodo.14048411}.

\section*{Code availability}
The reconstruction and selection of $e^+e^-$ collision events rely on the BESIII offline software system~\cite{Zou:2024pmc}. The training and inference of the GNN model use the open-source tool Weaver~\cite{weaver}, implemented with PyTorch~\cite{Paszke:2019xhz}. All algorithms used for data analysis and simulation are archived by the authors and are available on request to besiii-publications@ihep.ac.cn. The specific data analysis code is available at {\sc zenodo} repository \href{https://doi.org/10.5281/zenodo.14048411}{https://doi.org/10.5281/zenodo.14048411}.

\bibliography{main}

\clearpage
\section*{Acknowledgements}
The authors thank Huilin~Qu, Congqiao~Li, Sitian~Qian, Haiyong~Jiang and Jun~Xiao for suggestions on deep learning. The BESIII Collaboration thanks the staff of BEPCII and the IHEP computing center for their strong support. This work is supported in part by National Key R\&D Program of China under Contracts Nos. 2020YFA0406400, 2020YFA0406300, 2023YFA1606000; National Natural Science Foundation of China (NSFC) under Contracts Nos. 11635010, 11735014, 11835012, 11935015, 11935016, 11935018, 11961141012, 12025502, 12035009, 12035013, 12061131003, 12192260, 12192261, 12192262, 12192263, 12192264, 12192265, 12221005, 12225509, 12235017; the Chinese Academy of Sciences (CAS) Large-Scale Scientific Facility Program; the CAS Center for Excellence in Particle Physics (CCEPP); Joint Large-Scale Scientific Facility Funds of the NSFC and CAS under Contract No. U1832207; CAS Key Research Program of Frontier Sciences under Contracts Nos. QYZDJ-SSW-SLH003, QYZDJ-SSW-SLH040; 100 Talents Program of CAS; CAS Project for Young Scientists in Basic Research No. YSBR-117; The Institute of Nuclear and Particle Physics (INPAC) and Shanghai Key Laboratory for Particle Physics and Cosmology; European Union's Horizon 2020 research and innovation programme under Marie Sklodowska-Curie grant agreement under Contract No. 894790; German Research Foundation DFG under Contracts Nos. 455635585, Collaborative Research Center CRC 1044, FOR5327, GRK 2149; Istituto Nazionale di Fisica Nucleare, Italy; Ministry of Development of Turkey under Contract No. DPT2006K-120470; National Research Foundation of Korea under Contract No. NRF-2022R1A2C1092335; National Science and Technology fund of Mongolia; National Science Research and Innovation Fund (NSRF) via the Program Management Unit for Human Resources \& Institutional Development, Research and Innovation of Thailand under Contract No. B16F640076; Polish National Science Centre under Contract No. 2019/35/O/ST2/02907; The Swedish Research Council; U. S. Department of Energy under Contract No. DE-FG02-05ER41374.

\section*{Author Contributions}
The BESIII Collaboration (all contributing authors, as listed at the end of this manuscript) have contributed to the publication, being variously involved in the design and the construction of the detector, in writing software, calibrating sub-systems, operating the detector and acquiring data and finally analysing the processed data.

\section*{Competing Financial Interests}
The authors declare no competing interests.

\clearpage
\section*{Authors}
\begin{small}
\noindent M.~Ablikim$^{1}$, M.~N.~Achasov$^{4,c}$, P.~Adlarson$^{75}$, O.~Afedulidis$^{3}$, X.~C.~Ai$^{80}$, R.~Aliberti$^{35}$, A.~Amoroso$^{74A,74C}$, Q.~An$^{71,58,a}$, Y.~Bai$^{57}$, O.~Bakina$^{36}$
% ...
\\
\vspace{0.2cm}
(BESIII Collaboration)\\
\vspace{0.2cm} {\it
$^{1}$ Institute of High Energy Physics, Beijing 100049, People's Republic of China\\
% ...
$^{3}$ Bochum  Ruhr-University, D-44780 Bochum, Germany\\
$^{4}$ Budker Institute of Nuclear Physics SB RAS (BINP), Novosibirsk 630090, Russia\\
% ...
$^{35}$ Johannes Gutenberg University of Mainz, Johann-Joachim-Becher-Weg 45, D-55099 Mainz, Germany\\
$^{36}$ Joint Institute for Nuclear Research, 141980 Dubna, Moscow region, Russia\\
% ...
$^{57}$ Southeast University, Nanjing 211100, People's Republic of China\\
$^{58}$ State Key Laboratory of Particle Detection and Electronics, Beijing 100049, Hefei 230026, People's Republic of China\\
% ...
$^{71}$ University of Science and Technology of China, Hefei 230026, People's Republic of China\\
% ...
$^{74}$ University of Turin and INFN, (A)University of Turin, I-10125, Turin, Italy; (B)University of Eastern Piedmont, I-15121, Alessandria, Italy; (C)INFN, I-10125, Turin, Italy\\
$^{75}$ Uppsala University, Box 516, SE-75120 Uppsala, Sweden\\
% ...
$^{80}$ Zhengzhou University, Zhengzhou 450001, People's Republic of China\\

$^{a}$ Deceased\\
% ...
$^{c}$ Also at the Novosibirsk State University, Novosibirsk, 630090, Russia\\
% ...
}

\end{small}
\end{document}